\pdfoutput=1
\documentclass[conference]{IEEEtran}
\IEEEoverridecommandlockouts
\usepackage{cite}
\usepackage{amsmath,amssymb}
\usepackage{booktabs}
\usepackage{multirow}
\usepackage{graphicx}
\graphicspath{{Fig/}}
\usepackage[table]{xcolor}
\definecolor{rowhl}{gray}{0.92}
\usepackage{url}
\usepackage[hidelinks]{hyperref}
\usepackage{xurl}%
\usepackage{xspace}
\newcommand{\tool}{\textsc{CacheLint}\xspace}

\newcommand{\ying}[1]{#1}%

\usepackage[most]{tcolorbox}
\definecolor{rqaccent}{RGB}{30,64,175}
\definecolor{rqbg}{RGB}{239,244,255}
\newtcolorbox{rqbox}[1]{%
  enhanced, breakable, arc=2pt,
  colback=rqbg, colframe=rqaccent!80,
  boxrule=0.5pt,
  left=2.5mm, right=2.5mm, top=1.5mm, bottom=1.5mm,
  fonttitle=\bfseries\sffamily\small, coltitle=white,
  colbacktitle=rqaccent!90, titlerule=0pt,
  title={Answer to RQ#1},
  before skip=6pt, after skip=6pt,
}
\newcommand{\rqanswer}[2]{\par\medskip\begin{rqbox}{#1}\itshape #2\end{rqbox}\par\smallskip}
\definecolor{grpProv}{RGB}{198,40,40}
\definecolor{grpApp}{RGB}{21,101,192}
\definecolor{grpAcct}{RGB}{27,128,68}
\definecolor{grpImpl}{RGB}{124,92,30}
\definecolor{grpCross}{RGB}{142,124,195}
\definecolor{cSilent}{RGB}{180,83,9}
\definecolor{cCrash}{RGB}{185,28,28}
\definecolor{cCost}{RGB}{146,64,14}
\definecolor{cLatent}{RGB}{91,33,182}
\definecolor{cMetric}{RGB}{21,128,61}
\newcommand{\pill}[2]{\colorbox{#1!15}{\textcolor{#1!85!black}{\footnotesize\strut\,#2\,}}}
\newcommand{\sySilent}{\pill{cSilent}{Silent}}
\newcommand{\syCrash}{\pill{cCrash}{Crash}}

\newcommand{\syMetric}{\pill{cMetric}{Wrong report}}

\usepackage{listings}
\usepackage[altpo]{backnaur}%
\usepackage{algorithm}
\usepackage{algpseudocode}
\definecolor{diffadd}{RGB}{34,134,58}
\definecolor{diffdel}{RGB}{179,29,52}
\definecolor{kwblue}{RGB}{26,54,196}
\definecolor{strgreen}{RGB}{24,128,40}
\definecolor{apipurple}{RGB}{124,40,160}
\definecolor{linenumc}{gray}{0.55}
\lstdefinestyle{diff}{basicstyle=\scriptsize\ttfamily,breaklines=true,columns=fullflexible,%
  numbers=none,frame=single,rulecolor=\color[gray]{0.82},framesep=4pt,xleftmargin=4pt,%
  xrightmargin=2pt,backgroundcolor=\color[gray]{0.98},keepspaces=true,aboveskip=4pt,belowskip=2pt,%
  commentstyle=\itshape\color[gray]{0.45},%
  moredelim=[l][\color{diffdel}]{-},moredelim=[l][\color{diffadd}]{+}}
\tcbuselibrary{listings,skins}
\definecolor{lstframe}{RGB}{30,64,175}
\lstdefinestyle{diffbare}{language=Python,basicstyle=\scriptsize\ttfamily,%
  breaklines=true,columns=fullflexible,frame=none,keepspaces=true,%
  numbers=left,numberstyle=\tiny\color{linenumc},numbersep=7pt,stepnumber=1,%
  aboveskip=0pt,belowskip=0pt,%
  keywordstyle=\bfseries\color{kwblue},stringstyle=\color{strgreen},%
  commentstyle=\itshape\color[gray]{0.45},%
  emph={cache_control,supports_cache,setdefault,isoformat,append,registry,%
        startswith,ephemeral},emphstyle=\color{apipurple},%
  moredelim=*[l][\color{diffdel}]{-},moredelim=*[l][\color{diffadd}]{+}}
\newtcblisting[auto counter]{difflisting}[2][]{%
  enhanced, float=t, floatplacement=t,
  listing only, listing style=diffbare,
  colback=gray!2, colframe=lstframe!72!black, boxrule=0.6pt,
  arc=2.5pt, outer arc=2.5pt,
  top=3pt, bottom=3pt, left=7pt, right=5pt,
  fonttitle=\bfseries\footnotesize, coltitle=lstframe!88!black,
  colbacktitle=rqbg, titlerule=0pt, toptitle=2pt, bottomtitle=2pt,
  title={Listing~\thetcbcounter.\hspace*{0.4em}#2},
  label={#1}}
\newcommand{\nMergedPR}{622}

\newcommand{\nDefect}{120}

\newcommand{\nGroupA}{47}

\newcommand{\nReqSide}{97}
\newcommand{\nReadSide}{23}

\begin{document}
\title{Silent and Costly: Characterizing and Detecting Prompt-Caching Mis-Integration in LLM Agents}
\author{
\IEEEauthorblockN{Ying Zhang\textsuperscript{$\star$}}
\IEEEauthorblockA{\textit{Wake Forest University}\\
Winston-Salem, USA\\
ying.zhang@wfu.edu}%
\thanks{$\star$\,Equal contribution to this work.}
\and
\IEEEauthorblockN{Ruoqi Guo\textsuperscript{$\star$}}
\IEEEauthorblockA{\textit{Griffith University}\\
Gold Coast, Australia\\
ruoqi.guo@griffithuni.edu.au}
\and
\IEEEauthorblockN{Yi Liu\textsuperscript{\textsection}}%
\thanks{\textsection\,Corresponding author.}
\IEEEauthorblockA{\textit{Griffith University}\\
Gold Coast, Australia\\
yi.liu@griffith.edu.au}
\and
\IEEEauthorblockN{Gelei Deng}
\IEEEauthorblockA{\textit{Nanyang Technological University}\\
Singapore\\
gelei.deng@ntu.edu.sg}
\and
\IEEEauthorblockN{Yuekang Li}
\IEEEauthorblockA{\textit{University of New South Wales}\\
Sydney, Australia\\
yuekang.li@unsw.edu.au}
\and
\IEEEauthorblockN{Yepang Liu}
\IEEEauthorblockA{\textit{Southern University of Science and Technology}\\
Shenzhen, China\\
liuyp1@sustech.edu.cn}
\and
\IEEEauthorblockN{Leo Zhang}
\IEEEauthorblockA{\textit{Griffith University}\\
Gold Coast, Australia\\
leo.zhang@griffith.edu.au}
}

\maketitle

\begin{abstract}

Large language model (LLM) services process massive volumes of user requests, resulting in substantial inference costs. To reduce both cost and latency, providers have introduced prompt caching. However, correctly integrating prompt caching is non-trivial due to heterogeneous provider-specific APIs and continuously evolving caching semantics.
To understand prompt-caching mis-integration in downstream applications, we conduct the first empirical study of 120 prompt-caching-related pull requests from 10 popular open-source agent systems. Our findings show that prompt-caching mis-integrations arise in both request-side misconfiguration and read-side usage miscount. More importantly, most mis-integrations fail silently without causing observable runtime failures.

Motivated by these findings, we design \tool{}, which models a prompt-caching mis-integration as an invocation site together with the preconditions required for correct cache integration. \tool{} locates caching invocation sites with CodeQL and discharges each precondition either against a dated provider-capability oracle or with an LLM guard check. %
\tool{} achieves \ying{0.71} precision and 0.78 recall in identifying the \nDefect{} labeled mis-integrations across the 10 studied agent systems.
Our evaluation further demonstrates the effectiveness of \tool in real-world settings, identifying 32 mis-integrations across 571 agent projects \ying{with 0.86 precision}. These findings have received 21 developer confirmations and led to 18 upstream fixes, including one in Anthropic's official SDK.
\end{abstract}


\section{Introduction}
\label{sec:intro}

Large language model (LLM) services process billions of requests from end users every day, incurring considerable inference costs~\cite{projectdiscovery-caching}. Prior work~\cite{promptcache-gim} shows that a substantial fraction of prompts are identical, such as system prompts and few-shot examples used in multi-turn conversations. When such prompts are decomposed into multiple requests to LLM services, the same prefix is repeatedly sent across requests. Consequently, LLMs repeatedly reprocess these identical prompt portions, leading to redundant computation and substantial token costs. By mid-2025, enterprise-level LLM API costs were estimated to have reached \$8.4 billion~\cite{menlo-llm-2025}.

To address this inefficiency, LLM providers have introduced \textit{\textbf{prompt caching}}~\cite{anthropic-pc,openai-pc,gemini-cc}, which reduces redundant computation over identical prompt prefixes by reusing previously computed intermediate results. Specifically, user prompts are cached based on their token prefixes; a \textit{cache hit} occurs when the incoming request shares an exact token-prefix match with a previously cached prompt, starting from the first token (i.e., byte-for-byte equality). In this case, the cached computation is reused instead of recomputing the prefix. Otherwise, a \textit{cache miss} occurs, and the prompt is processed normally while its prefix may be inserted into the cache for future reuse. Any modification, insertion, or reordering invalidates the match from that point onward~\cite{anthropic-pc,brenndoerfer-caching}.

However, correctly integrating prompt caching is non-trivial because provider-specific configuration settings vary and the underlying mechanisms evolve. For example, Anthropic~\cite{anthropic-pc} and Amazon Bedrock~\cite{bedrock-pc} require applications to explicitly opt into caching and enforce limits on the number of cache entries per request, whereas OpenAI and Gemini apply caching automatically~\cite{openai-pc,gemini-cc}. Different providers adopt different caching enablement settings. 
Moreover, providers' prompt caching semantics can evolve over time. For instance, Bedrock introduced an opt-in one-hour cache lifetime tied to specific models~\cite{bedrock-ttl-1h}, while Anthropic later reduced the default cache duration from one hour to five minutes, which significantly affected hit rates in unchanged deployments~\cite{anthropic-ttl-change}.

\begin{figure*}
    \centering
    \includegraphics[width=0.95\linewidth]{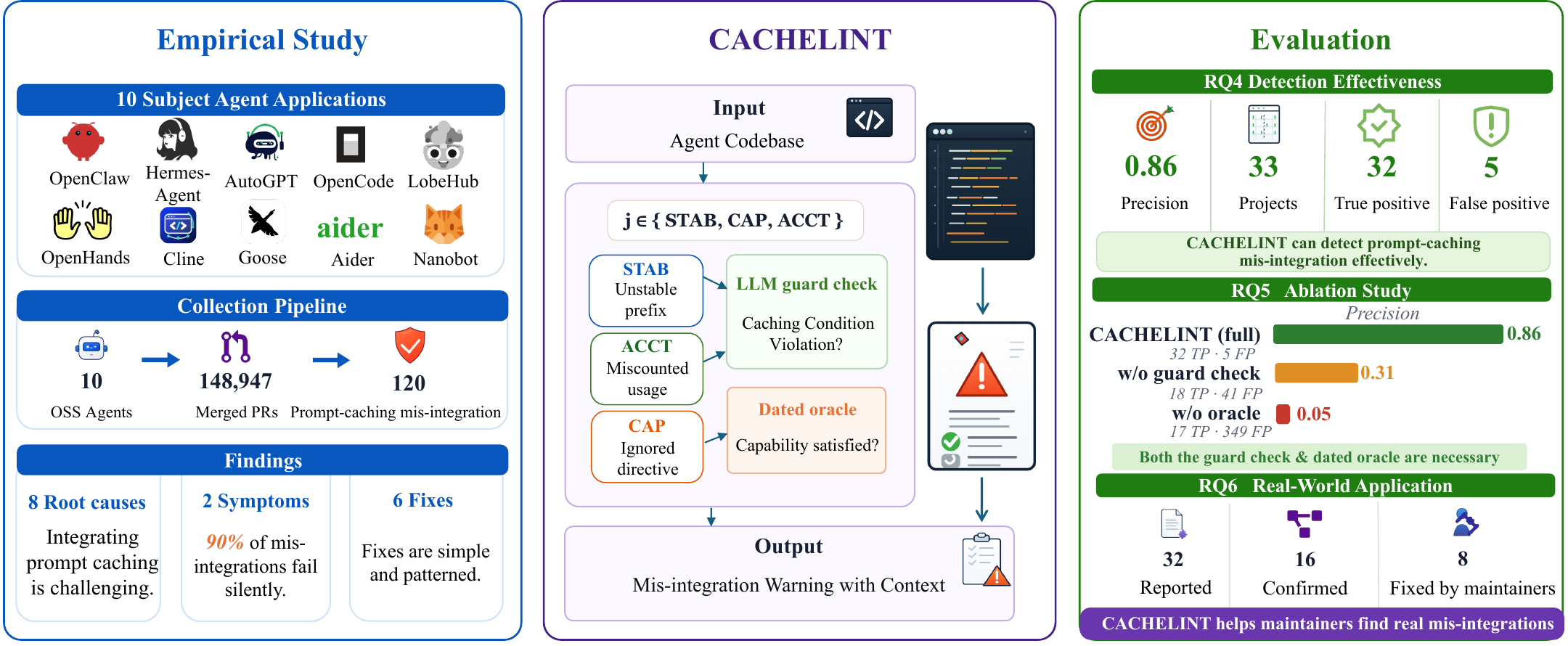}
    \vspace{-1em}
    \caption{The overview of our work.}
    \label{fig:overview}
\end{figure*}

Existing studies have investigated prompt defects and prompt caching performance, but have overlooked the integration issues of prompt caching from the application side. For instance, 
several existing studies~\cite{llmenginebugs-liu,llm-orchestration-bug-study,agentbugs-modern} investigate software defects that degrade the performance and reliability of LLM-integrated systems, including inference engines or agent orchestration frameworks. They do not consider prompt caching a distinct source of integration defects. Later, Tian et al.~\cite{tian2025promptdefects} identified prompt caching as a literature-derived subtype of performance and efficiency defects in a prompt defect taxonomy, while Lumer et al.~\cite{lumer2026dontbreakcache} evaluated prompt caching strategies for reducing latency and API cost. Neither of these studies investigates prompt cache misses caused by application-level integration.

In this paper, we conduct the \textit{first} empirical study of \textbf{\emph{prompt-caching mis-integration}} in open-source agent frameworks, analyzing application code that violates provider requirements for enabling, configuring, or accounting for prompt caching. As described in Fig.~\ref{fig:overview}, we identify 581 repositories through GitHub search and select the 10 most popular open-source agent applications for analysis. We then collect 148,947 merged pull requests (PRs) from these repositories and analyze their prompt-caching-related code. After inspection, we obtain \nDefect{} such defects. Our study explores the following research questions (RQs):
\begin{itemize}
    \item \textbf{RQ1 (Defect Type and Root Cause):} What are the common types of prompt caching defects in open-source agents, and what are their root causes?
    \item \textbf{RQ2 (Defect Symptom):} What are the common symptoms of prompt caching integration defects in open-source agents?
    \item \textbf{RQ3 (Defect Fix):} How do agent developers fix prompt caching integration defects in practice, and what are the common patterns?
\end{itemize}

Our study uncovers three key findings. First, prompt caching integration defects are concentrated on the request side. Of the \nDefect{} defects, \nReqSide{} corrupt outgoing requests, while the remaining \nReadSide{} preserve requests but misparse cache-usage metrics.
Interestingly, we find that the heterogeneous provider-specific caching mechanisms account for \nGroupA{} request-side defects, while inconsistent application-side specification of prompt prefixes (e.g., system prompts) is another major cause of cache misses.
Second, \textit{prompt-caching mis-integration is often silent.}  In 108 mis-integration cases, defects produce no errors because the model recomputes outputs without using cached tokens. The resulting cache misses manifest only as increased cost and latency in usage metrics.
Third, 
the mis-integration fixes are typically small and localized to request-construction code. 
These findings suggest that prompt-caching mis-integrations often preserve functional outputs and evade dynamic test oracles, making static analysis a well-suited detection approach.

In this paper, we develop \tool{}, a static analysis technique for detecting prompt-caching mis-integrations in agent applications. As shown in Fig.~\ref{fig:overview}, 
\tool pairs each prompt-caching invocation site with its required preconditions, derived from the fixing patterns of our empirical study.
Given an agent application codebase, \tool{} uses CodeQL~\cite{codeql} to identify prompt-caching API invocation sites and extract their surrounding code context. 
It then checks each associated precondition
by either comparing provider-capability preconditions against a dated capability oracle or prompting an LLM (\texttt{gpt-5.4-mini}) with the invocation-site context to assess prefix-stability and accounting preconditions. When a precondition check fails, \tool{} reports the corresponding invocation site and suggests a fix.

We evaluate \tool on the 10 studied projects, using \nDefect{} fixed defects as ground truth. \tool reports 132 warnings, of which 94 are true positives, achieving a precision of 0.71 and a recall of 0.78. To assess its generalizability, we further evaluate \tool on 571 real-world agent repositories, where it achieves a precision of 0.86 and uncovers 32 previously unknown prompt-caching mis-integrations. At the time of reporting, maintainers had confirmed 21 reports and fixed 18 defects. These results demonstrate the practical value of our empirical findings and \tool.

In summary, we make the following contributions: 
\begin{itemize}
\item 
We present the first empirical study of prompt-caching mis-integrations, characterizing their causes, symptoms, and fixes. We release our dataset to support future research~\cite{cachelint2026artifacts}.

\item We propose a prompt-caching integration model that captures common mis-integration patterns and generalizes developer practices for automated detection and repair.

\item We develop \tool{}, a static analysis-based detector for identifying prompt-caching mis-integrations. \tool{} achieves \ying{0.71 precision and 0.78 recall on the \nDefect{} labeled mis-integrations}. Furthermore, on 571 real-world agent applications, \tool uncovers 32 previously unknown defects \ying{at 0.86 precision}, resulting in 21 maintainer-confirmed reports and 18 fixes. 
\end{itemize}

\section{Background}
\label{sec:bg}

\subsection{Prompt Caching}
Prompt caching is an optimization technique that reduces LLM inference cost and latency by reusing the computation of previously processed prompt prefixes. Specifically, when a new prompt begins with a prefix that exactly matches a cached prefix, the provider reuses the cached computation instead of recomputing the prefix and charges cached tokens at a substantially reduced rate~\cite{anthropic-pc,openai-pc,gemini-cc}.

Prompt caching is effective for LLM agents because of frequent prompt reuse.
In each call, the agent constructs a prompt from a stable preamble, typically a system instruction, tool definitions, and retrieved or environment context, followed by the conversation history~\cite{claudecode-caching}. Across consecutive calls, these prompts share a long common prefix, while only the latest user message or model response changes. Without prompt caching, the agent repeatedly transmits and the LLM repeatedly processes thousands of identical prompt tokens. By reusing these stable prefixes, prompt caching can reduce cached input costs to roughly one-tenth of the standard price and decrease inference latency by up to 80\%~\cite{anthropic-pc,openai-pc,gemini-cc,projectdiscovery-caching}. Consequently, modern agent frameworks are increasingly designed to maximize cache reuse~\cite{reasonix}.

\subsection{Prompt Caching Integration} 
Prompt caching integration refers to how applications construct and configure prompts to satisfy provider-defined requirements for enabling caching behavior. When correctly integrated, prompt caching enables reuse of previously computed prompt prefixes, resulting in cache hits; otherwise, the system falls back to full recomputation.

The correct integration of prompt caching places two requirements on applications. \emph{First, the application must maintain prefix stability.} As providers perform byte-level matching from the first token, the prompt prefix must exactly match a previously cached prefix to guarantee cache hits. Even minor modifications to the prefix can invalidate cache reuse from that point onward~\cite{anthropic-pc,openai-pc,brenndoerfer-caching}.
\emph{Second, the application must correctly configure provider-exposed caching parameters.} For example, settings such as cache prefix length and cache lifetime determine whether a request is eligible for caching and whether cached entries can be reused. Insufficient prefix length or short cache windows may prevent reuse of previously computed results, leading to cache misses even when prompts are structurally identical.

\section{Motivating Example}
\label{sec:motiv}
We introduce a real prompt caching integration issue from PrivateGPT~\cite{Zylon_PrivateGPT_2023}. PrivateGPT is an open-source retrieval-augmented generation (RAG) framework~\cite{rag} that enables users to interact with private documents using an LLM through a context-aware question-answering interface. 

\begin{figure}[t]\centering
\includegraphics[width=0.95\columnwidth]{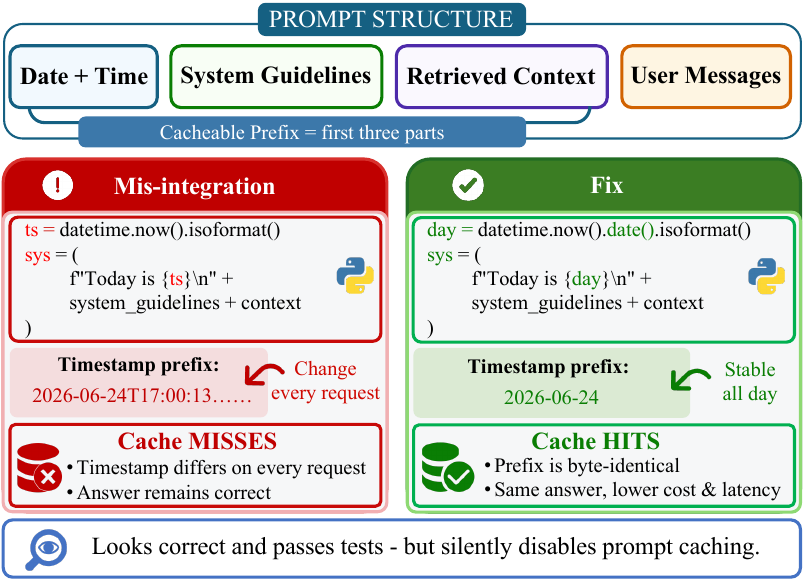}
\vspace{-1em}
\caption{A prompt-caching mis-integration in \texttt{zylon-ai/private-gpt}: a sub-second timestamp at the front of the system prompt breaks the cached prefix on every call; the fix coarsens it to day granularity.}
\label{fig:motiv}
\end{figure}

The developer attempted to enable the agent to answer queries about the current time while benefiting from prompt caching. 
The original implementation prepends \texttt{datetime.now().isoformat()} to the system prompt. Since cache hits require an exact match of the prompt prefix from the first token, the changing timestamp invalidates cache reuse for all subsequent stable content (left side of Fig.~\ref{fig:motiv}).
As a result, despite the provider correctly implementing prompt caching, the application fails to reuse cached prompt prefixes and repeatedly recomputes the stable guidelines and retrieved context.
Additionally, because the agent continues to produce correct outputs without raising any errors, this mis-integration remains silent until increased inference cost or a reduced cache-hit rate is observed.

To address this issue, the developer coarsens the timestamp to date granularity using \texttt{.date()}, which enables the prompt prefix to remain stable throughout the day while still providing accurate time information (right side of Fig.~\ref{fig:motiv}). 

We define such \emph{prompt-caching mis-integration} as \ying{application code that violates provider requirements for enabling, configuring, or accounting for prompt caching}. Such defects can prevent cache reuse, cause incorrect cache accounting, or even trigger runtime failures. Despite their practical impact, real-world reports show that many of these defects remain unnoticed in production until operational metrics, such as cache-hit rates or API costs, are examined~\cite{brenndoerfer-caching,hermes-13631}. This gap motivates our empirical study and the development of automated detection of prompt-caching mis-integration.

\section{Empirical Study Methodology}
This section presents the data collection and analysis methods we use to answer RQ1--RQ3.

\subsection{Dataset Collection}
\label{sec:dataset}

To study RQ1--RQ3, we automatically collect real-world agent projects from GitHub, including their source code, issues, and pull requests (PRs) related to prompt-caching defects.
We first adopt a keyword-based search strategy using three categories of prompt-caching implementation indicators: provider cache markers (\texttt{cache\_control}, \texttt{cachePoint}, \texttt{cachedContent}, \texttt{prompt\_cache\_key}), usage fields (\texttt{cached\_tokens}, \texttt{cache\_read}, \texttt{cache\_creation}), and the literal phrase ``prompt caching''. The search returned 581 repositories.

We then filter out irrelevant repositories and retain candidates that meet three selection criteria: (1) \emph{scope}--- %
general-purpose conversational or coding agent \emph{applications} for end users, excluding
inference engines, editors with an incidental agent, and single-purpose utilities; (2) \emph{active maintenance}---non-archived repositories with commits within three months of our collection snapshot (2026-05-10); (3) \emph{prompt-caching integration}---verified source-level prompt-caching implementations, excluding unrelated caches (e.g., HTTP, session, and key-value caches).
These criteria yield 581 candidate repositories. 

To bound the manual annotation effort, we rank the remaining repositories by GitHub stars and keep the top ten. These criteria yield the ten subjects in Table~\ref{tab:subjects}. They span TypeScript, Python, and Rust and cover the major providers: Anthropic, OpenAI, Gemini, Bedrock, and OpenRouter-style gateways. 

\begin{table}[t]
\centering
\scriptsize
\setlength{\tabcolsep}{3.5pt}
\renewcommand{\arraystretch}{1.15}
\caption{Top ten agent applications (\#PC = merged PRs that fix a prompt-caching defect).}\label{tab:subjects}
\begin{tabular}{@{}lrrr@{}}
\toprule
\textbf{Project} & \textbf{\#Star} & \textbf{\#PR} & \textbf{\#PC} \\
\midrule
OpenClaw~\cite{subj-openclaw}     & $379{,}764$ & $53{,}078$ & $39$ \\
Hermes-Agent~\cite{subj-hermes}   & $198{,}679$ & $37{,}707$ & $21$ \\
AutoGPT~\cite{subj-autogpt}       & $185{,}051$ & $8{,}235$  & $6$  \\
OpenCode~\cite{subj-opencode}     & $176{,}850$ & $13{,}884$ & $9$  \\
LobeHub~\cite{subj-lobehub}       & $78{,}924$  & $8{,}479$  & $1$  \\
OpenHands~\cite{subj-openhands}   & $77{,}888$  & $10{,}430$ & $11$ \\
Cline~\cite{subj-cline}           & $63{,}600$  & $5{,}897$  & $19$ \\
Goose~\cite{subj-goose}           & $49{,}956$  & $7{,}248$  & $7$  \\
Aider~\cite{subj-aider}           & $46{,}526$  & $1{,}047$  & $1$  \\
Nanobot~\cite{subj-nanobot}       & $44{,}499$  & $2{,}942$  & $6$  \\
\midrule
\textbf{Total} & - & $\mathbf{148{,}947}$ & $\mathbf{\nDefect}$ \\
\bottomrule
\end{tabular}
\end{table}

\textit{Identifying prompt-caching mis-integration.} Within these subjects, we identify each defect through the merged PR that fixes it. Defects without a merged fixing PR provide no root cause or fix strategy to analyze and are therefore excluded. We collect 148,947 merged PRs across the ten subjects and query the GitHub Search API using prompt-caching-related keywords, restricted to \texttt{type:pr is:merged created:<=2026-05-10}. This search returned \nMergedPR{} candidate merged PRs.

Because keyword matching is noisy, three authors manually screened all \nMergedPR{} candidates to separate genuine defects from false matches, reading each candidate's title, body, and, where needed, its diff. %
The first three authors independently screened all candidate PRs, achieving a Fleiss' $\kappa$ of 0.76. Disagreements were resolved through discussion. Screening retained \nDefect{} relevant prompt-caching PRs and discarded 502 false matches
involving unrelated caches, model additions, dependency updates, documentation, and generic token accounting. An independent second pass over all candidates identified no additional false matches among the retained PRs. Table~\ref{tab:subjects} summarizes the dataset used in our study.

\subsection{Data Analysis and Annotation}
\label{sec:annotation}
To understand prompt-caching integration defects and answer our research questions, we proceed in two steps. First, for each defect we (1) collect the fixing code, (2) recover the link between the defect report and the fixing code, and (3) collect and study the discussion thread. Second, three authors, all agent-application developers with at least two years of agent development and usage experience, apply open coding~\cite{open-coding}, an established qualitative research method, to classify each defect by its root cause, symptom, and fixing strategy. \ying{Each author coded all defects independently, and disagreements were resolved by consensus.}

\section{Empirical Study Results}
\label{sec:results}

\begin{table*}[t]\centering
\caption{Taxonomy of the \nDefect{} prompt-caching defects: two types and eight root causes, each with its count and share, mechanism, most common symptom, and most common fix. Bold rows give group subtotals.}
\label{tab:taxonomy}
\footnotesize
\setlength{\tabcolsep}{3pt}
\renewcommand{\arraystretch}{1.0}
\begin{tabular*}{\textwidth}{@{\extracolsep{\fill}}l l r r >{\raggedright\arraybackslash}p{6.3cm} l >{\raggedright\arraybackslash}p{2.7cm}@{}}
\toprule
\textbf{Type} & \textbf{Root cause} & \textbf{\#} & \textbf{\%} & \textbf{Description} & \textbf{Symptom} & \textbf{Fixing strategy} \\
\midrule
\multirow{10}{*}{\shortstack[l]{\textbf{Request side}\\{\footnotesize \nReqSide{} (81\%)}}}
 & \textbf{\textcolor{grpProv!85!black}{Provider compatibility}} & \textbf{47} & \textbf{39} &  &  &  \\
 & \quad Model-name gating & 37 & 31 & A hardcoded model-name check guards caching; a new model fails it. & \sySilent & Capability lookup \\
 & \quad Incompatible marker & 10 & 8 & The cache marker's format is one the provider rejects. & \syCrash & Marker guard \\
\cmidrule(l){2-7}
 & \textbf{\textcolor{grpApp!85!black}{Prefix stability}} & \textbf{41} & \textbf{34} &  &  &  \\
 & \quad Prefix reassembly & 15 & 13 & Each call rebuilds the prefix in a different order, breaking the byte-exact match. & \sySilent & Cache-boundary fix \\
 & \quad Volatile value & 13 & 11 & A per-request value (timestamp, UUID) is embedded in the cached prefix. & \sySilent & Cache-boundary fix \\
 & \quad Nondeterministic serialization & 10 & 8 & Identical content serializes to different bytes (e.g., unsorted map keys). & \sySilent & Prefix canonicalization \\
 & \quad Misplaced marker & 3 & 3 & The cache marker is placed on a changing message, not the stable prefix. & \sySilent & Marker guard \\
\cmidrule(l){2-7}
 & \textbf{\textcolor{grpCross!85!black}{Cross-turn mutation}} & \textbf{9} & \textbf{8} & Program state mutated between turns invalidates a prefix an earlier turn cached.   &  \sySilent & Lifecycle guard \\
\midrule
\multirow{2}{*}{\shortstack[l]{\textbf{\ying{Read side}}\\{\footnotesize \nReadSide{} (19\%)}}}
 & \textbf{\textcolor{grpAcct!85!black}{Usage miscount (read-side)}} & \textbf{23} & \textbf{19} & Caching works, but returned cache-token counts are summed or shown incorrectly. & \syMetric & Usage-counter fix  \\
\bottomrule
\end{tabular*}
\end{table*}

\subsection{RQ1: Defect Type and Root Causes}
\label{sec:rq1}

Table~\ref{tab:taxonomy} shows that \nReqSide{} defects (81\%) are on the request side, and the remaining \nReadSide{} (19\%) are on the read side.

\subsubsection{Request Side}
Request-side mis-integration occurs when an agent constructs or configures a request incorrectly, causing the provider to recompute content that could otherwise be served from the cache, thereby incurring the full inference cost, or causing the request to fail at runtime.

\begin{difflisting}[lst:gating]{Model-name gating, the largest cause (\texttt{cline\#3343}, simplified): caching keyed on a model name; the fix keys on a provider capability.}
  block = msg.setdefault("content", {})
- if model.id.startswith("claude") or "anthropic" in model.id:
+ if registry.supports_cache(model.id):
      block["cache_control"] = {"type": "ephemeral"}
\end{difflisting}

\textbf{Provider compatibility.} 
The agent adds a provider-specific caching configuration that is unsupported by the selected model or provider, causing the provider to ignore the configuration and leave prompt caching inactive.
Two root causes account for this failure. 

First, \emph{model-name gating} 
enables caching based on hardcoded model identifiers rather than provider capabilities. As a result, a model rename or the addition of a new model can silently disable caching.
Listing~\ref{lst:gating} shows a model-name-based mis-integration in \texttt{cline\#3343}. Specifically, an upstream rename changed a Gemini~2.5 model ID, but the application enabled caching only for models whose names began with \texttt{claude} or contained \texttt{anthropic}. As a result, the renamed model no longer passed the check, and caching was disabled. Similar failures occurred in \texttt{OpenHands\#8365}, where a \texttt{-latest} tag was removed; \texttt{hermes-agent\#17425}, which excluded MiniMax's cacheable models; and \texttt{OpenHands\#4704}, where an overly restrictive condition disabled caching altogether. 
Second, an \emph{incompatible marker} occurs when the agent attaches a provider-specific caching marker that the selected provider or model does not support. For example, 
\texttt{nanobot\#2687} attached Anthropic \texttt{cache\_control} markers to requests for every model accessed through OpenRouter. Because GPT, Gemini, and DeepSeek models do not support this marker, the requests resulted in 401 errors and a runtime crash.

\begin{difflisting}[lst:volatile]{Volatile value in the cached prefix (\texttt{goose\#2434}, simplified): a per-turn timestamp in the system prompt; the fix moves it out of the cached span.}
  system_prompt = [
-     f"Current time: {now.isoformat()}",
      "You are a coding agent. ...",
  ]
+ # per-turn time now rides the (uncached) user message
+ messages.append({"role": "user",
+     "content": f"[time] {now.isoformat()}"})
\end{difflisting}

\textbf{Prefix stability.} The agent's own prefix is not stable from one call to the next, so it never matches the prefix the previous call cached. Four root causes destabilize the prefix. A \emph{volatile value} embeds a per-request datum in the cached span: as shown in Listing~\ref{lst:volatile}, \texttt{goose\#2434} placed a live timestamp in the cached system prompt, changing the prefix on every call so the cache never hit. Similarly, \texttt{nanobot\#1115} did the same with a minute-level clock, and in each case the fix moves the volatile datum out of the cached span. \emph{Prefix reassembly} rebuilds the prefix in a changed order or shape each call, as in \texttt{hermes-agent\#754}, where a gateway reassembled the system prompt every turn and lifted cost by roughly 75\%. \emph{Nondeterministic serialization} emits different bytes for identical content, as in \texttt{openclaw\#64198}'s per-instance map order and \texttt{hermes-agent\#3991}'s random \texttt{uuid4} call-id. A \emph{misplaced marker} sits on a changing message rather than on the stable prefix.

\textbf{Cross-turn mutation.} A cross-turn mutation lets a stateful change between turns rewrite a fragment an earlier turn had already cached, so the next call writes the cache again instead of reading it. In \texttt{openclaw\#58038}, pruning history images after three turns rewrote the already-cached conversation prefix mid-session, so each compaction silently rebuilt the cache instead of hitting it.

\subsubsection{Read Side}
Read-side mis-integrations occur when caching works exactly as intended, but the agent mishandles the cache usage the provider reports back in the response. 

\textbf{Usage miscount.} The provider reports cache usage in its response, but the agent incorrectly parses, aggregates, prices, or displays the returned cache-token counts. In \texttt{cline\#10266}, the cache-creation tokens were hardcoded to zero in two provider handlers, so the usage readout reported no caching even as the cache kept hitting.

\rqanswer{1}{We observed eight root causes of prompt-caching mis-integrations. Provider compatibility and prefix stability account for the majority. Integrating prompt caching correctly across various providers is challenging.}

\subsection{RQ2: How Defects Manifest}
\label{sec:rq2}
Table~\ref{tab:taxonomy} shows that 108 of the \nDefect{} defects (90\%) remain silent. They raise no exception, no failed assertion, and no crash, and the model output stays correct while only caching degrades. We discuss the symptoms below.

Within these no-error defects, the vast majority, 86 (72\%), are silent waste: the cache stops hitting and the run continues unchanged, so the only trace is extra spend on a hosted API or added latency on the application, visible only on a usage meter. \texttt{cline\#3343} is typical, where a renamed model slipped past a caching gate and the bill climbed with no other sign. The remaining 22 (18\%) of the no-error defects produce wrong telemetry: caching works as intended, but the cache-token counts reported to the developer are parsed, summed, or displayed incorrectly, so the meter a developer would trust to confirm caching is itself wrong.

Only 12 of the \nDefect{} defects (10\%) crash, and they concentrate in incompatible markers (8 of the 12); the other four are cross-turn mutations (2), a misplaced marker (1), and a usage miscount (1). A crash is the one symptom an ordinary oracle catches, because it surfaces as an explicit error the first time the request runs: \texttt{OpenHands\#3689} hit Vertex's \texttt{Extra inputs not permitted} rejection on a \texttt{cache\_control} field, \texttt{hermes-agent\#1216} drew a 400 from a cache marker on an empty text block, and \texttt{AutoGPT\#12789} raised a \texttt{TypeError} on a null cache-token field. A single run or test exposes each, whereas the other 90\% give no such signal.

In summary, detecting prompt-caching mis-integration is hard precisely because 90\% of the defects fail silently: \texttt{cline\#3343} let a renamed model slip past a caching gate while the bill simply climbed, \texttt{goose\#2434} placed a live timestamp in the system prompt so the cache never hit, and \texttt{cline\#10266} hardcoded cache-creation tokens to zero so even the usage meter under-reported caching. None of them raises an exception or changes the model's answer, so crash- and assertion-based oracles miss the majority by construction.

\rqanswer{2}{Prompt-caching mis-integrations are overwhelmingly silent: 108 of \nDefect{} (90\%) raise no error and surface only as extra spend or added latency, which makes oracles hard to define for testing this class of defect.}

\subsection{RQ3: Fixing Strategies}
\label{sec:rq3}
We studied the \nDefect{} merged fixes together with the discussion in their linked reports to explore (1) the common patterns and (2) the complexity of the fixing patches.

\begin{figure}[t]\centering
\includegraphics[width=0.8\columnwidth]{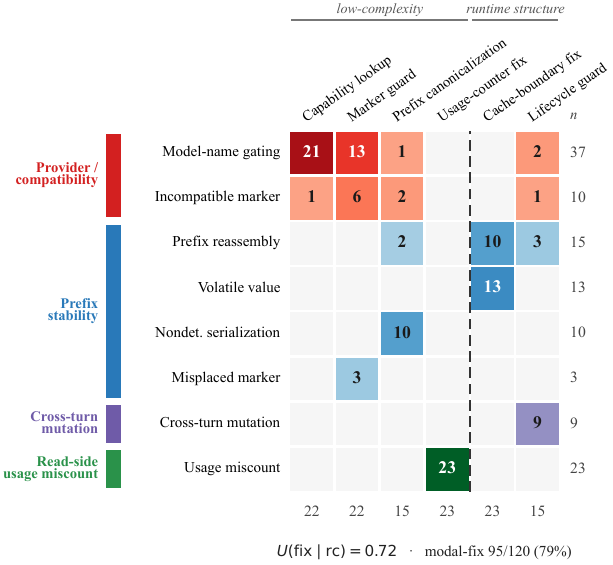}
\vspace{-1em}
\caption{Root cause vs.\ fixing strategy for the \nDefect{} defects. Each cell is a count, shaded by group and darker for larger counts.}
\label{fig:rq3}
\end{figure}

\textbf{Common fixing patterns.} As shown in Fig.~\ref{fig:rq3}, the eight root causes map to six recurring fixing strategies. Each strategy involves a relatively small, localized repair.
\emph{Capability lookup} resolves caching from a provider-capability check instead of a hardcoded model name; the \emph{marker guard} attaches the cache marker only in the shape and on the message the provider accepts; a \emph{cache-boundary fix} moves changing content out of the cached span (or rebuilds that span in a fixed order) so the prefix stays byte-identical; \emph{prefix canonicalization} serializes the prefix deterministically, e.g., by sorting map keys, so identical content always produces identical bytes; a \emph{lifecycle guard} stops a later turn from rewriting a prefix an earlier turn already cached; and a \emph{usage-counter fix} corrects the arithmetic that parses, sums, or displays the returned cache-token counts. 

Fixing strategies are largely predictable from the underlying root causes.
Simply assigning each root cause its most common fixing strategy correctly predicts 95 of the \nDefect{} fixes (79\%), with $U(\mathrm{fix}\mid\mathrm{rc})=0.72$. 
Five causes map to one strategy every time: usage miscounts to usage-counter fixes (23/23), volatile value to cache-boundary fixes (13/13), serialization to prefix canonicalization (10/10), cross-turn mutation to lifecycle guards (9/9), and misplaced markers to a marker guard (3/3). Only three cross-cutting causes spread their repairs: model-name gating splits between a capability lookup (21 of 37) and a marker guard (13), with a lifecycle guard (2) and prefix canonicalization (1) for the rest; an incompatible marker fans across four strategies, mostly a marker guard (6) with prefix canonicalization (2) and one each to a capability lookup and a lifecycle guard; and prefix reassembly divides among cache-boundary fixes (10), lifecycle guards (3), and prefix canonicalization (2). These three account for all 25 modal-fix errors, so the fix is largely, though not perfectly, a relabeling of the cause.

\textbf{Patch complexity.} Although the strategy varies by cause, the patches themselves are small. Four of the six strategies are static, local, or report-only: capability lookups (22, 18\%), marker guards (22, 18\%), prefix canonicalization (15, 13\%), and report-only usage-counter fixes (23, 19\%), and together they cover 82 of the \nDefect{} fixes (68\%); only cache-boundary fixes (23, 19\%) and lifecycle guards (15, 13\%) reach into runtime structure. The most common fix simply adds or corrects a single guard or the arithmetic on already-correct token counts.

\rqanswer{3}{Diagnosing a prompt-caching defect is the hard part: because it is mostly silent, the work lies in noticing it at all and tracing a cost or latency regression back to the request-building code. The fix itself is usually simple and patterned: most (68\%) are small, static, local edits that recur across projects as reusable templates.}

\section{Methodology}
\label{sec:detection}

Based on our empirical findings, we develop \tool{}, an automated technique for detecting prompt-caching mis-integrations in agent codebases. \tool{} models both prompt-caching-related APIs and the integration contexts associated with each class of mis-integration. Given an agent codebase, it performs static analysis to identify and report prompt-caching mis-integrations. We first present the underlying integration model and then describe the detection workflow.

\subsection{Prompt-Caching Integration Modeling}
\label{sec:model}
The \nDefect{} mis-integration cases arise from diverse root causes, but they share a common pattern: each either writes or reads the prompt cache incorrectly across providers. This regularity admits a uniform formalization. We model each pattern of prompt-caching mis-integration as a pair of a \emph{prompt-caching invocation site} $m$ and a \emph{caching-context precondition} $j$, written $P=\langle m, j\rangle$, where:

\begin{itemize}
  \item $m$ is a program point that sends content to, or reads an accounting field from, the cache. It is either a \emph{request-side} site that attaches a cache directive to a prefix span (a marker from $K=\{$\texttt{cache\_control}, \texttt{cachePoint}, \texttt{cachedContent}, a \texttt{ttl}/\texttt{cacheRetention} setting, \texttt{prompt\_cache\_key}$\}$) or a \emph{read-side} site that consumes a usage field from $U=\{$\texttt{cache\_read}, \texttt{cache\_creation}, \texttt{cached\_tokens}, \texttt{cacheReadInputTokens}$\}$.
  \item $j$ is a conjunction of three condition families, one per root-cause group in Table~\ref{tab:taxonomy} and generated by the grammar below: prefix stability (\textsc{stab}), provider capability (\textsc{cap}), and accounting safety (\textsc{acct}). The other two causes, cross-turn mutation and prefix reassembly, fall outside the grammar: each compares prefixes across different calls or turns, which a single program state cannot check.
\end{itemize}

{\scriptsize
\begin{bnf*}
\bnfprod{j}{\bnfpn{stab} \bnfor \bnfpn{cap} \bnfor \bnfpn{acct} \bnfor \bnfpn{j} \bnfsp \bnfts{and} \bnfsp \bnfpn{j}}\\
\bnfprod{stab}{\bnfts{volatile\_value\_reaches(span)} \; \bnftd{// clock/uuid/rng}}\\
\bnfmore{\bnfor \bnfts{nondet\_serialization(span)} \; \bnftd{// unsorted keys}}\\
\bnfmore{\bnfor \bnfts{marker\_on\_unstable(span)} \; \bnftd{// changing span}}\\
\bnfprod{cap}{\bnfts{predicate\_gates\_marker(model\_lit)} \bnfsp \bnfts{and}}\\
\bnfmore{\bnfsp\bnfsp \bnfts{dated\_oracle(model,prov,ttl,t)} \neq \mathrm{predicate}}\\
\bnfmore{\bnfor \bnfts{marker\_shape\_rejected(marker,prov,t)}}\\
\bnfprod{acct}{\bnfts{usage\_consumed\_without\_guard(U)} \; \bnftd{// int+None}}
\end{bnf*}
}

We define a mis-integration $M$ as an invocation reached on a path where a required caching precondition is not guaranteed to hold. Formally, $M = \langle m, j, \lnot\textsc{guard}\rangle$: an invocation site $m$, an unchecked conjunct $j\in\{\textsc{stab},\textsc{cap},\textsc{acct}\}$, and no preceding guard that discharges $j$ before control reaches $m$. A guard discharges $j$ if it checks the relevant condition, normalizes the inputs so that $j$ holds, or blocks the invocation when $j$ is false.
Among the three conjuncts, \textsc{cap} is unique because its truth depends on a time-varying external condition rather than the program itself. It may change long after the code is released without any modification to the implementation. For example, when Bedrock introduced a one-hour TTL, previously correct code silently became defective, whereas in other cases unchanged code became correct.

Consider \texttt{cline\#3343} (Listing~\ref{lst:gating}), which illustrates the model. At invocation site $m$, the code attaches \texttt{cache\_control} only when \texttt{model.id.startswith("claude")} holds. On provider \texttt{prov}, \texttt{gemini-2.5} is cacheable, but its identifier fails this gate, so \texttt{cache\_control} is dropped and the prefix is recomputed at full price. In the model, this mis-integration is the pair $\langle m,\,\textsc{cap}\rangle$ whose conjunct \texttt{dated\_oracle(gemini-2.5, prov, \_, t)} $\neq$ \texttt{predicate} is violated: the dated oracle marks \texttt{gemini-2.5} cacheable while the predicate admits only identifiers that begin with \texttt{claude}. Mined from fixed defects, the same pair flags other sites where the predicate excludes a model the oracle marks cacheable.

\begin{algorithm}[t]
\small
\caption{\tool{}: detecting a prompt-caching mis-integration at a cache-commit site.}
\label{alg:detect}
\begin{algorithmic}[1]
\Require source $A$; implemented pairs $P$; dated oracle $O$
\Ensure warnings $W$
\State $W \gets \emptyset$
\ForAll{$\langle m, j\rangle \in P$}
  \ForAll{$s \in \textproc{Locate}(A, m)$} \Comment{CodeQL / regex}
    \If{$j = \textsc{cap}$}
      \State $v \gets \textproc{OracleDiff}(\textproc{Assumption}(s),\, O)$ \Comment{dated, non-LLM}
    \Else \Comment{$\textsc{stab}$ or $\textsc{acct}$}
      \State $h \gets \textproc{LlmGuardCheck}(s,\, \textproc{Context}(A,s),\, j)$
      \State $v \gets h\ ?\ \textsc{ok} : \textsc{bug}$
    \EndIf
    \If{$v \neq \textsc{ok}$}
      \State $W \gets W \cup \{\langle s,\, j,\, v,\, \textproc{FixHint}(m,j)\rangle\}$
    \EndIf
  \EndFor
\EndFor
\State \Return $W$
\end{algorithmic}
\end{algorithm}

\begin{figure*}[t]
\centering
\includegraphics[width=0.8\textwidth]{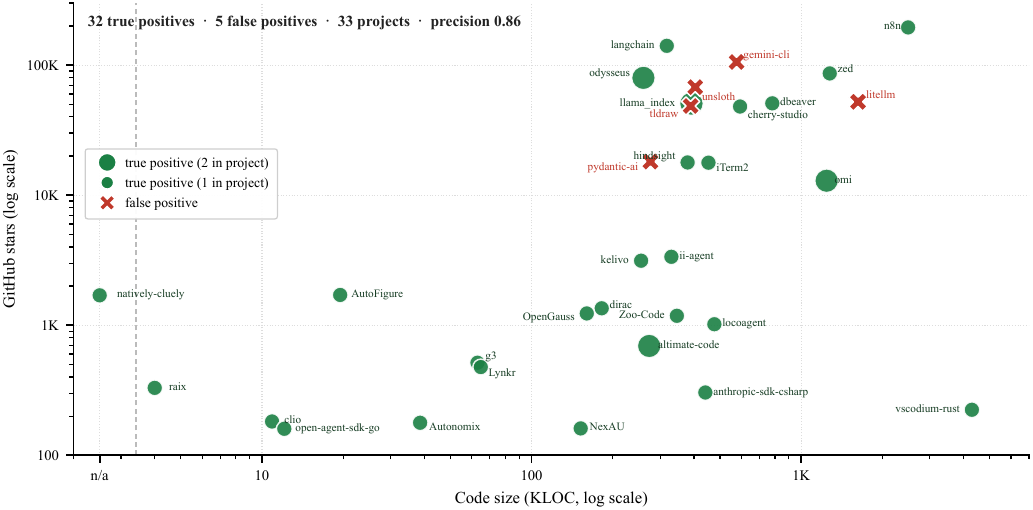}
\caption{\tool{}'s findings on 33 real-world projects, placed by code size (KLOC) and popularity (stars). }
\label{fig:rq4}
\end{figure*}

\subsection{Context-Guided Mis-Integration Identification}
\label{sec:tool}
Not every pair in Section~\ref{sec:model} is worth detecting. Each pair becomes a static rule, and a rule earns its place only when it fires reliably and recurs across projects, so \tool{} implements the anti-patterns that are (i) \emph{locatable}, in that the invocation site $m$ is a single, statically inspectable program point, and (ii) \emph{recurrent}, in that the pair generalizes across providers and apps rather than turning on one project's internals. This is the mechanizable subset of the taxonomy (Section~\ref{sec:rq4}); we leave out the multi-snapshot causes, cross-turn mutation and prefix reassembly, whose context a single program state cannot express.

Algorithm~\ref{alg:detect} takes an application's source, the implemented pairs, and the dated capability oracle $O$, and emits a warning for each unguarded invocation. For each pair it locates the invocation sites of the marker or meter, and for each site it discharges the pair's conjunct in one of two ways. A \textsc{cap} conjunct is settled \emph{off} the program: \tool{} extracts the code's capability assumption $\langle\textit{site},\textit{provider\ scope},\textit{capability},\textit{predicate}\rangle$ and diffs it against $O$, a dated external lookup the model never performs. A \textsc{stab} or \textsc{acct} conjunct is settled \emph{on} the program: \tool{} extracts the invocation site's surrounding context and asks an LLM whether a discharging barrier reaches $m$, such as a date-truncation or \texttt{sort\_keys} canonicalizer for \textsc{stab} or an \texttt{or 0} or accumulator guard for \textsc{acct}. This LLM guard check replaces the inter-procedural backward slice a static detector would use, which is brittle under the dynamic dispatch and reflection common in agent code. When no barrier discharges the conjunct, \tool{} emits the site, the verdict, and a suggested patch class drawn from RQ3's cause-to-fix map (a capability lookup for model-name gating, a cache-boundary fix for a volatile value, a usage-counter fix for a usage miscount), so every warning ships with the modal repair developers already apply.

For each \textsc{cap} assumption, \tool{} diffs the code's predicate against a provider-capability matrix carrying an explicit \texttt{last\_verified} date and a sourced behavior for each provider on an unsupported directive (\emph{reject}, \emph{ignore}, or \emph{unknown}). When the code treats a model as capable but the matrix does not (an over-claim), the verdict is a hard \textsc{bug} if the provider rejects the directive, \textsc{suspected} if its behavior is unknown, and benign if it silently ignores it; when the matrix marks the model as capable but the code excludes it (an under-claim), the verdict is a silent \textsc{latent} miss. Because the matrix is dated and external, the same one-hour-TTL code shape resolves to a \textsc{bug} on Bedrock and an \textsc{ok} on the Anthropic API, a distinction no static rule makes (Section~\ref{sec:rq5}).
\ying{Provider changes therefore reach \tool{} through two places: this matrix, and the marker list $K$ and usage-field list $U$ that the CodeQL queries behind \textproc{Locate} match. Every matrix entry is sourced from the provider's official documentation and stamped with the date of that check; when a provider changes its caching behavior, we update the affected entries and re-run \tool{}, leaving the CodeQL queries and the guard-check prompts unchanged.}

\subsection{Implementation}

The \textproc{Locate} step is a set of CodeQL queries~\cite{codeql}, one per anti-pattern and language (Python, JavaScript/TypeScript, Ruby, Go, C\#). The guard check is delegated to a frontier LLM, \texttt{gpt-5.4-mini}, prompted with the invocation site and its enclosing function and asked only whether a discharging barrier reaches $m$. As with prior static detectors~\cite{ficfinder}, this check is not sound: the LLM can overlook a real guard or posit a spurious one, so \tool{} may raise false warnings or miss defects. We release our source code at our website~\cite{cachelint2026artifacts}.

\section{Evaluation}
\label{sec:eval}
In this section, we evaluate \tool{} on real-world agent applications and investigate the following research questions:
\begin{itemize}
\item \textbf{RQ4 (Detection Effectiveness).} How effective is \tool{} at detecting prompt-caching mis-integrations?
\item \textbf{RQ5 (Ablation Study).} How does each component of \tool{} contribute to its overall effectiveness?
\item \textbf{RQ6 (Real-world Application).} Can \tool{} provide useful information to help developers identify prompt-caching mis-integrations?
\end{itemize}

\textbf{Datasets.} \ying{We evaluate \tool{} on two datasets, both drawn from the 581 repositories collected in Section~\ref{sec:dataset}.
\emph{Dataset~A} consists of the 10 agent applications from which we derived our detection model, together with the \nDefect{} mis-integrations labeled in them. Every defect in Dataset~A is linked to a merged fixing PR, so the dataset provides ground truth and supports measuring both precision and recall. Because the detection rules were mined from these same fixes, the recall measured on Dataset~A is an in-sample upper bound rather than an independent estimate.
\emph{Dataset~B} consists of the remaining 571 repositories. It is disjoint from Dataset~A, and therefore held out from rule derivation, and it is also broader, since it retains the agent SDKs, frameworks, and single-purpose tools that the empirical study excluded. No exhaustive defect labels exist for these repositories.
We apply \tool{} to the latest version of every repository in both datasets and manually inspect each warning in its source context to label it a true or false positive.}

\textbf{Evaluation metrics.} \ying{We label a warning a \emph{true positive} (TP) when the flagged site genuinely violates a prompt-caching precondition and a \emph{false positive} (FP) otherwise; on Dataset~A, a labeled defect that \tool{} misses is a \emph{false negative} (FN). We report $\mathit{precision}=\mathit{TP}/(\mathit{TP}+\mathit{FP})$ on both datasets, and $\mathit{recall}=\mathit{TP}/(\mathit{TP}+\mathit{FN})$ on Dataset~A only, since Dataset~B gives no denominator of known defects.}

All experiments were conducted on a MacBook Pro equipped with an Apple M5 Max processor and 48 GB of RAM. We present the experimental results below.

\subsection{RQ4: Detection Effectiveness}
\label{sec:rq4}
\ying{\noindent\textbf{Effectiveness on Dataset~A.} On Dataset~A, \tool{} reports 132 potential mis-integrations, comprising 94 TP and 38 FP, and fails to report 26 of the \nDefect{} labeled defects. It thus achieves a precision of 0.71 and a recall of 0.78.}

\ying{We manually inspected the code associated with the missed defects.  We found that most false negatives stem from defects that require reasoning across multiple invocation sites or turns, which are beyond the scope of the single-site model in \tool{}.  Of the \nDefect{} labeled defects, 96 occur at a single deterministic program point and are amenable to static inspection. These include model-name gating, usage miscounts, volatile cached-prefix values, nondeterministic serialization, misplaced markers, and incompatible markers. The remaining 24 defects require comparing prompts across calls or turns. They account for 24 of the 26 false negatives (15 involving prefix reassembly and 9 involving cross-turn mutation). The other two false negatives are in-scope volatile-value defects in which the changing values are introduced outside the enclosing function and thus fall beyond the context inspected by the guard check. Within its stated scope, \tool recovers 94 of the 96 in-scope defects, demonstrating near-complete coverage of defects amenable to single-site analysis.}

\ying{\noindent\textbf{Effectiveness on Dataset~B.}} \tool{} reports 37 potential mis-integrations across the 571 repositories of Dataset~B; Fig.~\ref{fig:rq4} shows 32 TP and 5 FP across 33 projects, a precision of 0.86. The detected defects cover a wide range of agent repositories, from flagship frameworks such as \texttt{n8n} (194,717 GitHub stars) to single-maintainer tools, with codebases ranging from 4 to 4,297 KLOC. Most findings fall into two categories: model-name gating and volatile values in the cached prefix. The remaining cases involve marker-budget and usage-miscount defects.

Two examples illustrate the results. In \texttt{dirac}, prompt caching was gated on a hard-coded Claude model in a \texttt{switch} statement, the same pattern as in our motivating \texttt{cline} example. The maintainers later merged a capability-based fix. In \texttt{n8n}, a millisecond timestamp was prepended to the cached system prompt, causing the prefix to change on every request and preventing cache hits. These results show that \tool{} can detect prompt-caching mis-integrations in both large and widely used agent projects.

The five false positives match the syntactic pattern of a mis-integration but are benign. Two occur because the caching requirement has already been satisfied: in \texttt{litellm}, cache-token accounting is safely guarded with \texttt{.get(\ldots{}, 0)}, and in \texttt{unsloth}, the system prompt uses a day-level \texttt{date().isoformat()}, which is stable enough for caching. In \texttt{gemini-cli}, a \texttt{Date.now()} value is used only as a request identifier and never appears in the cached prompt prefix. The remaining two arise from conservative matching: \texttt{tldraw} uses a provider feature flag that resembles model-name gating, and \texttt{pydantic-ai} contains a hard-coded model name only in test code. These cases are outside the scope of the current static rules and could be eliminated with more precise path-sensitive analysis.

We reproduced all 32 true positives. For the 5 crash defects, we triggered the runtime error. For the 27 silent defects, we inspected request/response pairs and the returned cache-usage fields (e.g., \texttt{cache\_read} and \texttt{cached\_tokens}).

\rqanswer{4}{\ying{\tool{} achieves a precision of 0.71 and a recall of 0.78 on the 10 popular agent applications, and a precision of 0.86 on 571 real-world agent applications.} The anti-patterns mined from fixed prompt-caching defects help \tool{} detect unknown mis-integrations in real-world agent projects, from flagship frameworks to single-maintainer tools.}

\subsection{RQ5: Ablation Study}
\label{sec:rq5}

\tool{} uses two procedures to discharge caching preconditions in Algorithm~\ref{alg:detect}. A \textsc{cap} conjunct is checked against the dated capability oracle, while \textsc{stab} and \textsc{acct} conjuncts are checked by the LLM-based guard check. 
To measure the contribution of each procedure, we construct two ablated variants and run them on Dataset~B. \ying{Each ablated variant removes one component and falls back to \texttt{gpt-5.4-mini} without the context that the component provides. \emph{\tool{} w/o guard check} judges every located \textsc{stab} or \textsc{acct} site with a generic API-misuse prompt instead of the targeted barrier question, for example whether a \texttt{date()} truncation or a \texttt{sort\_keys} canonicalizer reaches the site. \emph{\tool{} w/o oracle} judges every located \textsc{cap} site from the LLM's own knowledge.} Table~\ref{tab:ablation} reports the results.

The guard check is necessary for filtering benign \textsc{stab} and \textsc{acct} sites. Without it, precision drops from 0.86 to 0.31 because the detector reports many candidates whose caching preconditions have already been satisfied. For example, \texttt{unsloth} includes a timestamp in the system prompt, but the timestamp is coarsened to day granularity with \texttt{date().isoformat()}, making it stable enough for caching. The full \tool{} recognizes this guard and does not report the site, whereas \tool{} w/o guard check flags it as a volatile-value defect. Similarly, \texttt{litellm} safely accounts for cache tokens by guarding the lookup with \texttt{.get(\ldots{}, 0)} before summing the values; the ablated variant reports this benign site.

The dated capability oracle is also necessary. Without it, the model-name-gating query reports 349 candidate sites across the scanned repositories, most of which are benign branches in routers, registries, and tests. Precision drops to 0.05. The oracle distinguishes an actual capability mismatch from a harmless model gate. In \texttt{dirac}, for example, the oracle marks a renamed model as cacheable, while the code predicate excludes it; this disagreement exposes a real defect. In contrast, when a gated-out model is truly not cacheable, the code agrees with the oracle and the site is cleared. The oracle is also needed because these facts are time-sensitive: when asked about the 33 dated one-hour-TTL capability facts encoded in the matrix, \texttt{gpt-5.4-mini} answers only five confidently and correctly, and must abstain on models released after its training cutoff.

\ying{Both ablations also lower the number of true positives, from 32 to 18 without the guard check and to 17 without the oracle (Table~\ref{tab:ablation}). Without the component's context, the LLM tends to treat a prompt-caching mis-integration as an ordinary API call, so it misses real defects while still flagging many benign sites, which lowers both true positives and precision. The full \tool{} is stable across runs: over five runs, precision deviated between 0.0 and +0.02, and the deviation came from a single intermittently reported false positive.}

\rqanswer{5}{Both the LLM-based guard check and the dated capability oracle are necessary for \tool{}'s effectiveness. The guard check filters benign \textsc{stab} and \textsc{acct} sites, while the oracle filters benign \textsc{cap} sites. Removing either component sharply reduces precision.}

\begin{table}[t]
\centering
\footnotesize
\caption{Ablation of \tool{}.}
\label{tab:ablation}
\begin{tabular}{@{}lrrr@{}}
\toprule
\textbf{Configuration} & \textbf{TP} & \textbf{FP} & \textbf{Precision} \\
\midrule
\tool{} (full)        & 32 & 5   & 0.86 \\
\quad w/o guard check & 18 & 41  & 0.31 \\
\quad w/o oracle      & 17 & 349 & 0.05 \\
\bottomrule
\end{tabular}
\end{table}

\subsection{RQ6: Real-World Application}
\label{sec:rq6}

To evaluate the practical value of \tool{}, we reported the 32 true-positive findings from RQ4 to project maintainers and tracked their responses. \ying{At the time of writing, 21 reports have been confirmed and 18 have been fixed; 11 are awaiting a response and none has been rejected. The confirmations include widely used projects such as \texttt{AutoGPT}, \texttt{n8n}, and \texttt{Zed}.} The confirmed cases span provider SDKs, large frameworks, and single-maintainer tools. We discuss three representative examples.

\noindent\textbf{Usage miscount in the Anthropic SDK.}
\texttt{anthropic-sdk-csharp} is Anthropic's official .NET SDK. The API returns cache-creation and cache-read token counts as optional usage fields, but the SDK deserialized these fields into nullable properties and consumed them without a null check. When a response omits the fields, clients cannot reliably account for cache usage or cache savings. We reported the issue, and the maintainers fixed it.

\noindent\textbf{Model gating in \texttt{altimate-code}.}
\texttt{altimate-code} attached \texttt{cache\_control} markers only for Anthropic-native calls, using a hard-coded set of model paths. Identical requests routed through OpenRouter- or Bedrock-style gateways were therefore left unmarked and could not use prompt caching. This is the same request-side gating pattern as in the motivating \texttt{cline} case. We reported the issue, and the maintainers fixed it by replacing the model-list check with a capability-based check.

\noindent\textbf{Volatile prefix in \texttt{iTerm2}.} \texttt{iTerm2} is a widely used macOS terminal. Its assistant integration placed a request-varying value near the beginning of the cached system prompt. This changed the cached prefix on every call and prevented cache hits. We reported the issue, and the maintainer fixed it by moving the volatile value out of the cached region.

\rqanswer{6}{\tool{} can provide useful information to help developers diagnose and fix prompt-caching mis-integrations.}

\section{Threats to Validity}
\label{sec:threats-detector}

\noindent\textbf{Construct and internal validity.}
Our results depend on how defects are identified, labeled, and validated. To reduce noise from keyword search, three authors independently screened all candidate PRs and identified \nDefect{} genuine prompt-caching defects. Three experienced raters also independently coded all defects and resolved disagreements by consensus. To reduce evaluation bias, we inspected each \tool{} warning in source context and reported findings upstream, where maintainers confirmed \ying{21} reports and fixed \ying{18}. LLM judgments and provider capabilities may change over time. We mitigate this by fixing the prompts and model version, using a dated capability snapshot, and evaluating on 571 repositories. \ying{Recall on the 571 repositories is not measured because the total number of prompt-caching defects there is unknown; the 0.78 recall on the labeled corpus is an in-sample proxy, since the same corpus informed the detection rules.}

\noindent\textbf{External validity.}
Our taxonomy is derived from the ten most-starred multi-provider agent applications on GitHub and may not generalize to closed-source or non-GitHub projects. We mitigate this by evaluating \tool{} on 571 disjoint real-world repositories. \tool{} covers 96 (\(80\%\)) of the defects in our dataset; multi-snapshot defects remain out of scope. \ying{The empirical study was restricted to end-user agent applications and excluded SDKs and frameworks, which may bias the taxonomy toward application code; however, the 571-repository evaluation set includes SDKs, and the confirmed defect in Anthropic's official .NET SDK (Section~\ref{sec:rq6}) shows that the same patterns occur in provider-maintained code. \tool{} targets mis-integrations that involve explicit cache markers or parameters, including OpenAI's \texttt{prompt\_cache\_key} and Gemini's \texttt{cachedContent}. When a provider enables caching automatically on the model side, request-side defects are observable only through the response, for example as consistently zero cached tokens; such implicit caching is outside \tool{}'s current scope and is left for future work.} As provider caching semantics evolve, our results reflect the 2026-05-10 capability snapshot. \ying{The taxonomy is separated from provider-specific API details; as caching APIs evolve, the oracle entries and detection rules are updated from providers' published documentation, as described in Section~\ref{sec:tool}.}

\section{Related Work}

\emph{Compatibility, fragmentation, and API misuse.} Provider-specific caching behavior is related to platform compatibility and API-misuse research. FicFinder detects Android compatibility issues through API-context pairs~\cite{ficfinder,ficfinder-tse}, while CiD studies API evolution~\cite{cid}. Prompt caching introduces an additional dimension: application-side request invariants such as prefix stability and cache-marker accounting. Like API-misuse detectors~\cite{apimisuse-amann}, \tool{} checks whether program behavior satisfies an external API contract. Rather than relying on usage patterns alone, it combines data-flow analysis with provider knowledge. \ying{Prompt-caching mis-integration also resembles a configuration error. Studies of misconfiguration show that such errors can cause production failures and often arise from unchecked assumptions about the deployment environment or software version~\cite{yin2011config,xu2013misconfig,xu2015configsurvey,zhang2014config}. Model-name gates and cache-marker shapes are assumptions of this kind, written into code rather than into configuration files, and \tool{} checks them against a dated external oracle.}

\emph{Empirical bug studies and silent defects.} Repository-mining studies have characterized faults in deep-learning software~\cite{dlbugs-zhang,dltaxonomy-humbatova,islam-fse}, LLM inference engines~\cite{llmenginebugs-liu}, and agent frameworks~\cite{llm-orchestration-bug-study,agentbugs-modern}. We follow the same open-coding methodology~\cite{open-coding} to analyze 120 prompt-caching fixes. Prompt-caching defects belong to the broader class of silent performance bugs: they preserve functional correctness while increasing cost or latency, escaping traditional testing and crash-based oracles~\cite{perfbugs-jin,perfjs-selakovic,silentbugs-tambon}.

\emph{LLM-assisted program analysis.} Recent analyses combine LLM reasoning with static verification. InferROI infers resource intentions and validates them with control-flow analysis~\cite{inferroi}; IRIS infers taint specifications and grounds them with CodeQL~\cite{iris-li}. \tool{} adopts the same paradigm for prompt caching: CodeQL identifies candidate invocation sites~\cite{codeql}, an LLM guard check verifies prefix stability and usage accounting, and a dated capability oracle determines whether provider-specific caching assumptions remain valid.

\emph{Prompt caching and LLM serving.} Prior systems optimize prompt caching inside the serving engine through KV-cache management~\cite{vllm-pagedattention}, cross-request prefix reuse~\cite{sglang-radixattention}, and modular prompt reuse~\cite{promptcache-gim}. We instead study failures introduced by application code, where unstable prefixes or unsupported cache markers violate the engine's caching preconditions before a request reaches the provider. \ying{Closer to our topic, Tian et al.~\cite{tian2025promptdefects} list prompt caching as one category in a taxonomy of prompt defects, and Lumer et al.~\cite{lumer2026dontbreakcache} measure how caching strategies trade off cost against latency on long-horizon agentic tasks. Both treat caching as a defect category to catalog or as a strategy to tune; we instead characterize the defects that arise when applications integrate caching incorrectly and build a detector that flags them in code.}

\section{Conclusion}
\label{sec:conclusion}

In this paper, we presented the first empirical study of prompt-caching
mis-integrations in LLM agent applications. We investigated \nDefect{} real
defects from 10 open-source agent systems to characterize their root causes,
symptoms, and fixes. We found that these defects arise on both the request and
read sides, mostly fail silently while preserving model output, and are
typically repaired by small, localized edits. Guided by these findings, we
proposed a prompt-caching integration model that pairs each caching invocation
site with the preconditions for its correct use, and built \tool{}, a static
analysis technique that discharges each precondition against a dated
provider-capability oracle or an LLM guard check. On 571 real-world
repositories, \tool{} detected 32 previously unknown mis-integrations at 0.86 precision,
yielding \ying{21} developer confirmations and \ying{18} upstream fixes. In the future,
we plan to extend \tool{}'s oracle to more providers and to automatically
generate and validate fixes.

\section*{Acknowledgment}
\ying{We thank the anonymous reviewers for their constructive feedback and guidance, which helped us improve this paper.}

\section*{Data Availability}
We release the dataset and artifacts used in this work at our website~\cite{cachelint2026artifacts}.

\clearpage
\bibliographystyle{IEEEtran}
\bibliography{refs}

\end{document}